\documentclass{article}
\usepackage{spconf,amsmath}
\usepackage{amsthm}
\usepackage{amsfonts}
\usepackage{amssymb}
\usepackage{mathrsfs}
\usepackage{mathtools}
\usepackage[english]{babel}
\usepackage{float}
\usepackage[toc,page]{appendix}
\usepackage[pdftex]{graphicx}
\usepackage{epstopdf}
\usepackage{amsmath}
\usepackage{color}
\usepackage{multirow}
\usepackage{graphicx}
\usepackage{IEEEtrantools}
\usepackage{bm}
\usepackage{balance}
\usepackage{microtype}
\usepackage[caption=false,font=footnotesize]{subfig}
\usepackage{cite}
\usepackage{comment}
\allowdisplaybreaks

\numberwithin{equation}{section}

\newtheorem{theorem}{Theorem}[section]
\makeatletter
\newcommand{\settheoremtag}[1]{% \settheoremtag{<tag>}
  \let\oldthetheorem\thetheorem% Store \thetheorem
  \renewcommand{\thetheorem}{#1}% Redefine it to a fixed value
  \g@addto@macro\endtheorem{% At \end{theorem}, ...
    \addtocounter{theorem}{-1}% ...restore theorem counter value and...
    \global\let\thetheorem\oldthetheorem}% ...restore \thetheorem
  }
\makeatother

\usepackage[switch]{lineno}
\usepackage[named]{algo}
\usepackage[noend]{algpseudocode}
\usepackage{algorithm}

\newcommand{\suml}{\sum_{\ell=0}^{L-1}}
\newcommand{\sump}{\sum_{p=0}^{P-1}}
\newcommand{\sumk}{\sum_{k=0}^{K-1}}
\newcommand{\sumq}{\sum_{q=0}^{Q-1}}

\newcommand{\bsym}{\boldsymbol}

\newcommand{\en}[1]{\left[#1\right]}
\newcommand{\alpharl}{[\bsym{\alpha}_r]_\ell}
\newcommand{\taurl}{[\bsym{\bar{\tau}}_r]_\ell}
\newcommand{\nurl}{[\bsym{\bar{\nu}}_r]_\ell}
\newcommand{\alphacq}{[\bsym{\alpha}_c]_q}
\newcommand{\taucq}{[\bsym{\bar{\tau}}_c]_q}
\newcommand{\nucq}{[\bsym{\bar{\nu}}_c]_q}
\DeclareMathOperator*{\minimize}{minimize }

\newtheorem{proposition}[theorem]{Proposition}

\ninept

\newcommand{\dddag}{%
  \mathbin{\vbox{\offinterlineskip\ialign{%
    \hfil##\hfil\cr
    \small{$\dagger$}\cr
    \noalign{\kern-0.6ex}
    \small{$\ddagger$}\cr
}}}}

\title{Joint Radar-Communications Processing from a Dual-Blind Deconvolution Perspective} 
	
\name{Edwin Vargas$^{\dag}$, Kumar Vijay Mishra$^{\ddag}$, Roman Jacome$^{\dag}$, Brian M. Sadler$^{\ddag}$, and Henry Arguello$^{\dag}$}
\address{$^{\dag}$Universidad Industrial de Santander, Bucaramanga, Colombia, 680002\\ 
$^{\ddag}$United States CCDC Army Research Laboratory, Adelphi, MD 20783 USA
}

\begin{document}
% Reduce spacing above and below equations
\setlength{\abovedisplayskip}{3pt}
\setlength{\belowdisplayskip}{3pt}

\maketitle

\begin{abstract}
We consider a general spectral coexistence scenario, wherein the channels and transmit signals of both radar and communications systems are unknown at the receiver. In this \textit{dual-blind deconvolution} (DBD) problem, a common receiver admits the multi-carrier wireless communications signal that is overlaid with the radar signal reflected-off multiple targets. When the radar receiver is not collocated with the transmitter, such as in passive or multistatic radars, the transmitted signal is also unknown apart from the target parameters. Similarly, apart from the transmitted messages, the communications channel may also be unknown in dynamic environments such as vehicular networks. As a result, the estimation of unknown target and communications parameters in a DBD scenario is highly challenging. In this work, we exploit the sparsity of the channel to solve DBD by casting it as an atomic norm minimization problem. Our theoretical analyses and numerical experiments demonstrate perfect recovery of continuous-valued range-time and Doppler velocities of multiple targets as well as delay-Doppler communications channel parameters using uniformly-spaced time samples in the dual-blind receiver.
\end{abstract}

\begin{keywords}
Atomic norm, dual-blind deconvolution, channel estimation, joint radar-communications, passive sensing.
\end{keywords}
%\vspace{-8pt}
\section{Introduction}
\label{sec:intro}
With the advent of new wireless communications systems and novel radar technologies, electromagnetic spectrum has become a contested resource \cite{mishra2019toward}. This has led to the development of various system engineering and signal processing approaches for an optimal spectrum sharing performance \cite{paul2016survey}. In general, spectrum-sharing technologies follow three major approaches: co-design \cite{liu2020co}, cooperation \cite{bicua2018radar}, and coexistence \cite{wu2021resource}. While co-design requires designing new systems and waveforms to efficiently utilize the spectrum \cite{duggal2020doppler}, spectral cooperation requires exchange of additional information between radar and communications to enhance their respective performances \cite{paul2016survey}. The coexistence approach, on the other hand, does not mandate any additional hardware redesign or new waveforms. However, among all three approaches, separating the radar and communications signals at the coexistence receiver is the most challenging because of lack of any degrees-of-freedom in distinguishing the two signals. In this paper, we focus on the coexistence problem.

In general, coexistence systems employ different radar and communications waveforms and separate receivers, wherein the management of interference from different radio systems is key to retrieving useful information \cite{bicua2018radar}. When the received radar signal reflected-off from the target is overlaid with communications messages occupying the same bandwidth, knowledge of respective waveforms is useful in designing matched filters to extract the two signals \cite{paul2016survey}. Usually, the radar signal is known and the goal of the radar receiver is to extract unknown target parameters from the received signal. In a typical communications receiver, the channel is known and the unknown transmit message is of interest in the receive processing. However, these assumptions do not extend to a general scenario. For example, in a passive \cite{sedighi2021localization} or multistatic \cite{dokhanchi2019mmwave} radar system, the receiver may not have information about the radar transmit waveform. Similarly, in communications over dynamic channels such as millimeter-wave \cite{mishra2019toward} or terahertz-band \cite{elbir2021terahertz}, the channel coherence times are very short. As a result, the channel state information changes rapidly and the channel cannot be considered perfectly known. 

In this paper, we consider this general spectral coexistence scenario, where both radar and communications channels and the respective transmit signals are unknown to the common receiver. Accordingly, we model the extraction of all four of these quantities as a \textit{dual-blind deconvolution} (DBD). The formulation is an extension of the 
\textit{blind deconvolution} --- a longstanding problem that occurs in a variety of engineering and scientific applications, such as
astronomy, communication, image deblurring, system identification and optics --- wherein two unknown signals are estimated from an observation of their convolution~\cite{jefferies1993restoration,ayers1988iterative,abed1997blind}. This problem is ill-posed and, usually, structural constraints on signals are imposed to obtain algorithms with provable performance guarantees. The underlying ideas in these techniques are based on compressed sensing and low-rank matrix recovery, wherein signals lie in the low-dimensional random subspace and/or in high signal-to-noise ratio (SNR) regime \cite{lee2016blind,li2019rapid,ahmed2013blind,kuo2019geometry}. 

Previously, \cite{farshchian2016dual} studied the \textit{dual deconvolution} problem, where the radar transmit signal and communications channel were known. In this paper, unlike prior works, we examine the overlaid radar-communications signal as an ill-posed DBD problem. Our approach toward this more challenging problem is inspired by some recent works \cite{chi2016guaranteed,yang2016super} that have analyzed the basic blind deconvolution for off-the-grid sparse scenarios. The radar and communications channels are usually sparse \cite{mishra2018sub}, more so at higher frequency bands, and their parameters are continuous-valued \cite{mishra2015spectral}. Hence, sparse reconstruction in the off-the-grid or continuous parameter domain through techniques based on atomic norm minimization (ANM) \cite{off_the_grid} are appropriate for our application. While ANM has been extended to higher dimensions \cite{xu2014precise} and multiple parameters (e.g. delay-Doppler) \cite{heckel2016super}, prior works have not dealt with mixed radar-communications signal structures. We formulate an ANM-based recovery of both continuous-valued target parameters (range-time and Doppler velocities) as well as delay-Doppler communications channel estimates from a DBD receiver signal. We assume both channels are sparse, radar transmits a train of pulses and communications is a multi-carrier signal. Numerical experiments validate our approach through perfect recovery.

The rest of the paper is organized as follows. In the next section, we introduce the system model and DBD problem. We describe our ANM-based semi-definite program (SDP) to solve DBD in Section~\ref{sec:formulation}. We validate our model and methods in Section~\ref{sec:results} through numerical experiments. We conclude in Section~\ref{sec:summ}. 
% Notations (write as shown in the following example)
Throughout the paper, we reserve boldface lowercase, boldface uppercase, and calligraphic letters for vectors, matrices, and index sets, respectively. We denote the transpose, conjugate, Hermitian, and trace by $(\cdot)^T$, $(\cdot)^*$, $(\cdot)^H$, and $\text{Tr}(\cdot)$, respectively. The identity matrix of size $N\times N$ is $\mathbf{I}_N$. $||\cdot||_p$ is the $\ell_p$ norm. For notational convenience, the variables with subindex $r$ refer to the signals and parameters related to the radar system, while those with subindex $c$ refer to the communications system.
%\vspace{-8pt}
\section{System Model}
Consider a continuous-time linear system model that receives an overlaid radar-communications signal
\begin{align}
y(t) = x_r(t)*h_r(t)+x_c(t)*h_c(t),
\end{align}
where $x_r(t)$ is a train of $P$ transmitted pulses $s(t)$ with a pulse repetition interval (PRI) $T$; $x_c(t)$ is the transmitted communications signal composed of $P$ messages $\mathbf{g}_p$ with symbol duration $T$; and $h_r(t)$ and $h_c(t)$ are, respectively, the radar and communications channels modeled as a delay-Doppler channel with attenuation $\alpha$, time delay $\tau$ and shifting frequencies $\nu$. Assume $L$ targets/sources and $Q$ propagation paths for the radar and communications channels, respectively. Then, the set of vectors $\{{\bsym{\alpha}}_r \in \mathbb{C}^{L}, \bar{\bsym{\tau}}_r\in \mathbb{R}^{L}, \bar{\bsym{\nu}}_r \in \mathbb{R}^{L}\} $ and $\{{\bsym{\alpha}}_c \in \mathbb{C}^{Q}, \bar{\bsym{\tau}}_c \in \mathbb{R}^{Q}, \bar{\bsym{\nu}}_c  \in \mathbb{R}^{Q}\}$ contain the parameters of all $L$ targets and $Q$ paths, respectively. The communications messages are modulated by orthogonal frequency-division multiplexing (OFDM) waveform with $K$ sub-carriers, each separated by $\Delta f$. 

 The communications messages and radar pulses are transmitted in the same PRI. Then, the continuous-time received signal in the coherent processing interval (CPI) comprising $P$ pulse, i.e. $t\in[0,(P-1)T]$, is 
\begin{flalign}
    {y}(t) &= \sum_{p=0}^{P-1}\suml\alpharl s(t-pT-[\bar{\bsym{\tau}}_r]_\ell)e^{-\mathrm{j}2\pi\nurl pT}\nonumber\\&+\sum_{p=0}^{P-1}\sumq\sumk \alphacq [\mathbf{g}_p]_k e^{\mathrm{j}2\pi k\Delta f (t - pT - \taucq)}e^{-\mathrm{j}2\pi \nucq pT}.\nonumber
\end{flalign}
Expressing the received signal as  ${y}(t)= \sump \tilde{y}_p(t)$, our measurements are determined in terms of shifted signals $y_p(t) = \tilde{y}_p(t+ pT)$, such that the signals $\tilde{y}_p(t+ p\tau)$ are time-aligned with $y_0(t)$. Therefore, the signal $y_0(t)$ and the shifted signals $y_p(t)$ contain the same set of parameters. Take the continuous-time Fourier Transform (CTFT) of $y_p(t)$ in $f\in[-\frac{B}{2},\frac{B}{2}]$, with $p = 0,\dots, P-1$ and uniformly sample at $f_n= \frac{Bn}{M}=n\Delta f$, with $n = -N,\dots,N$, $M = 2N+1$. 

Assume $M = K$, i.e. sample in frequency domain at the OFDM separation frequency $\Delta f$ \cite{zheng2017super} to obtain
\begin{align}
    Y_p(f_n) = &\suml[\bsym{\alpha}_r]_\ell S(f_n)e^{-\mathrm{j}2\pi n \Delta f\taurl}e^{-\mathrm{j}2\pi\nurl pT}\nonumber\\
    &+\sumq [\bsym{\alpha}_c]_q e^{-\mathrm{j}2\pi n \Delta f\taucq }e^{-\mathrm{j}2\pi[\bar{\bsym{\nu}}_c]_{q}pT}[\mathbf{g}_p]_{n+N},
    \label{ft_2}
\end{align}
where $S(f)$ is the Fourier transform of $s(t)$. Concatenating $Y_p(f_n)$ in the vector $\mathbf{y}_p$, i.e., $\en{\mathbf{y}_p}_{n+N} = Y_p(f_n)$ for $n=-N,\cdots,N$, the discrete values of the Fourier transform in \eqref{ft_2} are
\begin{align}
    [\mathbf{y}_p]_{n+N} = &\suml[\bsym{\alpha}_r]_\ell [\mathbf{s}]_{n+N} e^{-\mathrm{j}2\pi(n[{\bsym{\tau}}_r]_\ell+ p[{\bsym{\nu}}_r]_\ell )}\nonumber\\&+\sumq [\bsym{\alpha}_c]_q[\mathbf{g}_p]_{n+N} e^{-\mathrm{j}2\pi(n[{\bsym{\tau}}_c]_\ell+ p[{\bsym{\nu}}_c]_\ell )},
    \label{eq:y_p}
\end{align}
where $[{\bsym{\tau}}_r]_\ell = \frac{[\bar{\bsym{\tau}}_r]_\ell}{T}$, $[\bsym{\nu}_r]_\ell = \frac{[\bar{\bsym{\nu}}_r]_\ell}{\Delta f}$, $[\bsym{\tau}_c]_\ell = \frac{[\bar{\bsym{\tau}}_c]_\ell}{T}$, $[\bsym{\nu}_c]_\ell = \frac{[\bar{\bsym{\nu}}_c]_\ell}{\Delta f} \in [0,1)$ are normalized sets of target parameters and  $\en{\mathbf{s}}_{n+N} = S(f_n)$. Define the vector $\mathbf{y} = \en{\mathbf{y}_0^T,\cdots,\mathbf{y}_{P-1}^T}^T$. Then, all samples of the $P$ received blocks of information are
\begin{align}
    &[\mathbf{y}]_{v} =\suml[\bsym{\alpha}_r]_\ell [\mathbf{s}]_{n+N} e^{-\mathrm{j}2\pi(n[{\bsym{\tau}}_r]_\ell+ p[{\bsym{\nu}}_r]_\ell )}\label{eq:y_v}\\&+\sumq [\bsym{\alpha}_c]_q[\mathbf{g}]_v e^{-\mathrm{j}2\pi(n[{\bsym{\tau}}_c]_q+ p[{\bsym{\nu}}_c]_q )},\;v=0,\cdots,MP-1,\nonumber
\end{align}
where $v = n +N+Np $ is the sequence index with $n = -N,\dots,N$ and $p = 0,\dots,P-1$, and $\mathbf{g} = [\mathbf{g}_0^T,...,\mathbf{g}_{P-1}^T]$. Note that $\en{\mathbf{y}}_v = \en{\mathbf{y}_p}_{n+N} $ and $\en{\mathbf{g}}_v = \en{\mathbf{g}_p}_{n+N}$. 

Our goal is to estimate the set of parameters $\bsym{\alpha}_r,\bsym{\tau}_r, \bsym{\nu}_r,\bsym{\alpha}_c,\bsym{\tau}_c$ and $\bsym{\nu}_c$ when the radar pulses $\mathbf{s}$ and communications symbols $\mathbf{g}$ are also unknown. To this end, we exploit the sparsity of the channels in our ANM formulation. We first employ the lifting trick \cite{chi2016guaranteed} that consists on representing the unknown signals in a low-dimensional sub-space, which follows the isotropy and incoherence properties \cite{candes2011probabilistic}. This implies spectral flatness over the represented signals. Highly spectrum-efficient waveforms such as OFDM \cite{ofdm_flat} or certain modulating Gaussian radar waveforms \cite{pralon2021analysis} satisfy this property. 

We represent the pulse signal $\mathbf{s}$ and the set of symbols $\mathbf{g}_p$ for $p=0,\cdots,P-1$ as $\mathbf{s} = \mathbf{B}\mathbf{u}$, where $\mathbf{B} = \en{\mathbf{b}_{-N},\dots, \mathbf{b}_{N}}^H\text{, }\mathbf{b}_n\in \mathbb{C}^J$, $\mathbf{g}_p = \mathbf{D}_p\mathbf{v}_p$, $\mathbf{D}_p = \en{\en{\mathbf{D}_p}_{-N},\cdots,\en{\mathbf{D}_p}_{N}}^H$, $\en{\mathbf{D}_p}_n \in \mathbb{C}^{J}$, and $\en{\mathbf{D}_p}_n$ denotes the $n$-th column of matrix $\mathbf{D}_p$. Define the full set of symbols $ \mathbf{g}$ as $\mathbf{g} = \mathbf{D}\mathbf{v}$, $\mathbf{D} =\en{\mathbf{d}_{0},\cdots,\mathbf{d}_{MP-1}}^H =\text{blockdiag}\en{\mathbf{D}_0^H,\cdots,\mathbf{D}_{P-1}^H}$, where $\textrm{blockdiag}\en{\mathbf{X}_1,\cdots,\mathbf{X}_n}$ is a block diagonal matrix with matrices $\mathbf{X}_1,\cdots, \mathbf{X}_n$. The matrices $\mathbf{B}\in\mathbb{C}^{M\times J}$ and $\mathbf{D}\in\mathbb{C}^{MP\times PJ}$ are known representation basis for the pulses and symbols and $\mathbf{u} \in \mathbb{C}^J$ and  $\mathbf{v}=[\mathbf{v}_0^T,\cdots,\mathbf{v}_P^T]^T \in \mathbb{C}^{PJ}$ are the corresponding coefficients vector with $J\ll N$. Using this representation, \eqref{eq:y_v} becomes:       
\begin{align}
    [\mathbf{y}]_{v} = &\suml[\bsym{\alpha}_r]_\ell \mathbf{b}_n^H\mathbf{u}e^{-\mathrm{j}2\pi(n[\bsym{\tau}_r]_\ell+p[\bsym{\nu}_r]_\ell  )}\nonumber\\&+\sumq [\bsym{\alpha}_c]_q \mathbf{d}_{v}^H\mathbf{v}e^{-\mathrm{j}2\pi (n[\bsym{\tau}_c]_q+p[\bsym{\nu}_c]_q )}.\label{signal}
\end{align}
Denote the channel vectors $\mathbf{h}_r = \suml[\bsym{\alpha}_r]_\ell\mathbf{a}(\mathbf{r}_\ell)$ and $\mathbf{h}_c = \suml[\bsym{\alpha}_c]_q\mathbf{a}(\mathbf{c}_q)$ where $\mathbf{r}_\ell = [[\bsym{\tau}_r]_\ell,[\bsym{\nu}_r]_\ell], \mathbf{c}_q = [[\bsym{\tau}_c]_q,[\bsym{\nu}_c]_q]$ are the parameter-tuples, and the vector $\mathbf{a}(\mathbf{r})$ is 
\begin{align}
    \mathbf{a}([\tau,\nu]) = \big[&e^{\mathrm{j}2\pi(\tau (-N)+\nu (0))},\dots,e^{\mathrm{j}2\pi(\tau (N)+\nu(0))},\nonumber\\&\dots,e^{\mathrm{j}2\pi(\tau (N)+\nu(P-1))}\big] \in\mathbb{C}^{MP}.
\end{align}
Then, \eqref{signal} becomes
\begin{align}
    [\mathbf{y}]_{v} = \mathbf{h}_r^H \mathbf{e}_{v} \mathbf{b}_n^H \mathbf{u} + \mathbf{h}_c^H\mathbf{e}_{v}\mathbf{d}_{v}^H\mathbf{v}\label{signal_2}.
\end{align}
Denote the matrices $\mathbf{Z}_r = \mathbf{u}\mathbf{h}_r^H \in \mathbb{C}^{J\times MP}$,  $\mathbf{Z}_c =\mathbf{v}\mathbf{h}_c^H \in \mathbb{C}^{PJ\times MP}$, $\mathbf{G}_v = \mathbf{e}_{v}\mathbf{b}_n^H \in \mathbb{C}^{MP\times J}$ and $\mathbf{A}_v = \mathbf{e}_{v}\mathbf{d}_{v}^H \in \mathbb{C}^{MP\times PJ}$, where $\mathbf{e}_{v}$ is the $v$-th canonical vector of $\mathbb{R}^{MP}$. Therefore, we  formulate \eqref{signal_2}  as $
        \mathbf{y} = \aleph_r(\mathbf{Z}_r) + \aleph_c(\mathbf{Z}_c)
$
where the linear operators   $\aleph_r: \mathbb{C}^{J\times MP}\rightarrow \mathbb{C}^{MP}$ and $\aleph_c: \mathbb{C}^{PJ\times MP}\rightarrow \mathbb{C}^{MP}$ are  
\begin{align}
    &[\aleph_r(\mathbf{Z}_r)]_{v} = \text{Tr}(\mathbf{G}_v\mathbf{Z}_r), \quad [\aleph_c(\mathbf{Z}_c)]_{v} = \text{Tr}(\mathbf{A}_v\mathbf{Z}_c).
\end{align}
Finally, we write the measured vector as
\begin{equation}
    \mathbf{y} = \aleph_r(\mathbf{Z}_r) + \aleph  _c(\mathbf{Z}_c).
    \label{eq:received_signal}
\end{equation}
Note that the matrices $\mathbf{Z}_r$ and $\mathbf{Z}_c$ contain all the unknown variables that we want to estimate. %In the following section, we leverage the sparsity of channels to solve this problem.
%\vspace{-8pt}
\section{Dual-Blind Deconvolution Algorithm} 
\label{sec:formulation}
The radar and communications channels are often characterized by a few continuous-valued parameters. Leveraging the sparse nature of these channels, we use ANM framework \cite{off_the_grid} for super-resolved estimations of continuous-valued channel parameters. %In this work, we exploit the same sparsity prior to finding an overlaid receiver signal's radar and communications channel. 
For the overlaid radar-communications signal, we formulate the parameter recovery as the minimization of two atomic norms, each corresponding to the radar and communications signal trails. Define the sets of atoms for the radar and communications signals as, respectively,
\begin{align}
    &\mathcal{A}_r = \Big\{\mathbf{u}\mathbf{a}(\mathbf{r})^H: \mathbf{r}\in[0,1)^2,||\mathbf{u}||_2 = 1   \Big\}
    \\&\mathcal{A}_c = \Big\{\mathbf{v}\mathbf{a}(\mathbf{c})^H: \mathbf{c}\in[0,1)^2,||\mathbf{v}||_2 = 1   \Big\}.
    \label{eq:atomic_sets}
\end{align}
The corresponding atomic norms are
\begin{align}
    &||\mathbf{Z}_r||_{\mathcal{A}_r} = \inf_{\stackrel{\alpharl \in \mathbb{C}, \boldsymbol{r}_\ell \in [0,1]^2}{||\mathbf{u}||_2 = 1}} \Bigg\{\sum_\ell |\alpharl| \Big| \mathbf{Z}_r = \sum_\ell \alpharl\mathbf{u}\mathbf{a}(\mathbf{r}_\ell)^H\Bigg\}\nonumber\\
    &||\mathbf{Z}_c||_{\mathcal{A}_c} = \inf_{\stackrel{\alphacq \in \mathbb{C}, \boldsymbol{c}_q \in [0,1]^2}{||\mathbf{v}||_2 = 1}} \Bigg\{\sum_q |\alphacq| \Big| \mathbf{Z}_c = \sum_q \alphacq\mathbf{v}\mathbf{a}(\mathbf{c}_q)^H\Bigg\}\nonumber.
\end{align}
Consequently, our proposed ANM problem is
\begin{align}
    &\minimize_{\mathbf{Z}_r,\mathbf{Z}_c} ||\mathbf{Z}_r||_{\mathcal{A}_r} +||\mathbf{Z}_c||_{\mathcal{A}_c}\nonumber\\
    &\text{subject to }     \mathbf{y} = \aleph_r(\mathbf{Z}_r) + \aleph_c(\mathbf{Z}_c).
    \label{eq:primal_problem}
\end{align}
%\subsection{Dual problem} 

In order to formulate the semidefinite program (SDP) of the above-mentioned ANM, we find its dual problem \cite{mishra2015spectral,xu2014precise} via standard Lagrangian analysis. The Lagrangian function of \eqref{eq:primal_problem} is 
\begin{align}
    \mathcal{L}(\mathbf{Z}_r,\mathbf{Z}_c,\mathbf{q})=||\mathbf{Z}_r||_{\mathcal{A}_r} +||\mathbf{Z}_c||_{\mathcal{A}_c}+  \langle\mathbf{q,y}-\aleph_r(\mathbf{Z}_r) - \aleph_c(\mathbf{Z}_c)\rangle
\end{align}
where $\mathbf{q}$ is the dual variable. The dual function $g(\mathbf{q})$ is obtained from the Lagrangian function as 
%\small
\begin{flalign}
    &g(\mathbf{q})\nonumber\\
    &=\inf_{\mathbf{Z}_r,\mathbf{Z}_c}\mathcal{L}(\mathbf{Z}_r,\mathbf{Z}_c,\mathbf{q})\nonumber\\
    &=\langle\mathbf{q,y}\rangle  \label{eq:dual}\\
    &\;\;+ \inf_{\mathbf{Z}_r,\mathbf{Z}_c} (||\mathbf{Z}_r||_{\mathcal{A}_r} +||\mathbf{Z}_c||_{\mathcal{A}_c})-\langle\mathbf{q},\aleph_r(\mathbf{Z}_r)\rangle-\langle\mathbf{q}, \aleph_c(\mathbf{Z}_c)\rangle)\nonumber\\
    &=\langle\mathbf{q,y}\rangle\nonumber \\
    &\;\;- \sup_{\mathbf{Z}_r} \left(\langle\aleph^*_r(\mathbf{q}),\mathbf{Z}_r \rangle-||\mathbf{Z}_r||_{\mathcal{A}_r}\right)-\sup_{\mathbf{Z}_c} \left(\langle\aleph^*_c(\mathbf{q}),\mathbf{Z}_c \rangle-||\mathbf{Z}_c||_{\mathcal{A}_c}\right).\nonumber 
\end{flalign}
%\normalsize
The supremum values in \eqref{eq:dual} correspond to the convex conjugate function of the atomic norms $||\mathbf{Z}_r||_{\mathcal{A}_r}$ and $||\mathbf{Z}_c||_{\mathcal{A}_c}$. For the atomic norm, $f=||\cdot||_\mathcal{A}$ the conjugate function is the indicator function of the dual norm unit ball
\[
  f^*(\mathbf{Z}) =
  \begin{cases}
                                   0 & \text{if } ||\mathbf{Z}||_{\mathcal{A}}^*\leq 1,\\
                                   \infty & \text{otherwise,}  \\
  \end{cases}
\]
where the dual norm is defined as
$
||\mathbf{Z}||_{\mathcal{A}}^* = \sup_{||\mathbf{U}||_{\mathcal{A}}\leq 1} \langle \mathbf{U},\mathbf{Z} \rangle
$.
Based on the dual function, the dual optimization problem of \eqref{eq:primal_problem} is
\begin{align}
    &\underset{\mathbf{q}}{\textrm{maximize}}\langle\mathbf{q,y}\rangle_{\mathbb{R}}\nonumber\\
    &\text{subject to } \Vert\aleph_r^\star(\mathbf{q})\Vert^\star_{\mathcal{A}_r}\leq1, \nonumber\\
    &\;\;\;\;\;\;\;\;\;\;\;\;\;\;\; \Vert \aleph_c^\star(\mathbf{q})\Vert^\star_{\mathcal{A}_c}\leq1, 
\label{eq:dual_problem_op}    
\end{align}
where $\aleph_r^\star: \mathbb{C}^{MP}\rightarrow\mathbb{C}^{J \times MP}$ and $\aleph_c^\star: \mathbb{C}^{MP}\rightarrow\mathbb{C}^{PJ \times MK}$  are adjoint operators of $\aleph_r$ and $\aleph_c$ respectively, i.e. $\aleph_r^\star(\boldsymbol{q}) = \sum_{p=0}^{P-1}\sum_{n=-N}^{N}[\mathbf{q}]_{v}\mathbf{G}_v^H,\aleph_c^\star(\mathbf{q}) = \sum_{p=0}^{P-1}\sum_{n=-N}^{N}[\mathbf{q}]_{v}\mathbf{A}_v^H $. 

The SDP of this dual problem is found using the trigonometric vector polynomials
\begin{align}
    &\mathbf{f}_r(\mathbf{r}) =  \sum_{p=0}^{P-1}\sum_{n=-N}^{N}[\mathbf{q}]_{v} \mathbf{G}_v^H \mathbf{a}(\mathbf{r}) \in \mathbb{C}^{J},
    \label{eq:poly_r}\\
    &\mathbf{f}_c(\mathbf{c})=\sum_{p=0}^{P-1}\sum_{n=-N}^{N}[\mathbf{q}]_{v}\mathbf{A}_v^H \mathbf{a}(\mathbf{c}) \in \mathbb{C}^{JP}.
    \label{eq:poly_c}
\end{align}
In particular, the SDP relaxation is achieved through the Bounded Real Lemma \cite{dumitrescu2007positive} to convert the constraints on \eqref{eq:dual_problem_op} into linear matrix inequalities and using the fact that polynomials \eqref{eq:poly_r} and \eqref{eq:poly_c} are parameterized by positive definite matrices. The relaxation of \eqref{eq:dual_problem_op} leads to the following equivalent SDP problem  
\begin{align}
    &\underset{\mathbf{q,Q}}{\textrm{maximize}}\quad \langle\mathbf{q,y}\rangle_{\mathbb{R}}\nonumber\\
    &\text{subject to }\mathbf{Q}\succeq 0,\nonumber\\&\hphantom{\text{subject to }}  
    \begin{bmatrix}
        \mathbf{Q} & \hat{\mathbf{Q}}_r^H \\
        \hat{\mathbf{Q}}_r & \mathbf{I}_J 
        \end{bmatrix}
    \succeq0,\nonumber\\&\hphantom{\text{subject to }} 
    \begin{bmatrix}
        \mathbf{Q} & \hat{\mathbf{Q}}_c^H \\
        \hat{\mathbf{Q}}_c & \mathbf{I}_{JP} 
        \end{bmatrix}\succeq 0,
    \nonumber\\&\hphantom{\text{subject to }}
    \text{Tr}(\boldsymbol{\Theta}_\mathbf{n}\boldsymbol{Q}) = \delta_{\mathbf{n}},\label{dual_opt}
\end{align}
where $\hat{\mathbf{Q}}_r =  \sum_{p=0}^{P-1}\sum_{n=-N}^{N}[\mathbf{q}]_{v}\mathbf{G}_v^H \in \mathbb{C}^{MP\times J}$ and $\hat{\mathbf{Q}}_c = \sum_{p=0}^{P-1}\sum_{n=-N}^{N} [\mathbf{q}]_{v} \mathbf{A}_v^H \in \mathbb{C}^{MP\times PJ}$ are the coefficients of two-dimensional (2-D) trigonometric polynomials, the matrix $\boldsymbol{\Theta}_\mathbf{n} $, $\mathbf{n} = [n_1,n_2]$ for the 2-D case is $\boldsymbol{\Theta}_\mathbf{n} = \boldsymbol{\Theta}_{n_2} \otimes \boldsymbol{\Theta}_{n_1}$, where $\bsym{\Theta}_n$ is the Toeplitz matrix with ones in the $n$-th diagonal with $0<n_1<m_1$ and $-m_2<n_2<m_2$. Here, we define $m_1 = P-1$ and $m_2=N-1$. Finally, $\delta_\mathbf{n} = 1 $ if $\mathbf{n} = [0,0]$ and $0$ otherwise.  This SDP formulation is solved by employing off-the-shelf solvers. 

Based on the strong duality implied by Slater's conditions, the following proposition states the conditions for exact recovery of the radar and communications channels parameters.  
\begin{proposition}
Let $\mathbf{y}$ be as defined in \eqref{eq:received_signal} and the atomic sets $\mathcal{A}_r$ and $\mathcal{A}_c$ as defined in \eqref{eq:atomic_sets}. Let $\mathcal{R} = \{\mathbf{r}_\ell\}_{\ell=0}^{L-1}$ and $\mathcal{C} = \{\mathbf{c}_q\}_{q=0}^{Q-1}$ and the solution of \eqref{eq:primal_problem} be $\hat{\mathbf{Z}}_r$ and $\hat{\mathbf{Z}}_c$. Then, $\hat{\mathbf{Z}}_r={\mathbf{Z}}_r$ and $\hat{\mathbf{Z}}_c={\mathbf{Z}}_c$ are the primal optimal solutions of \eqref{eq:primal_problem} if the following condition is satisfied:
There exist two 2-D trigonometric polynomials %in $\boldsymbol{\tau}_r$, $\bsym{\nu}_r$ and $\boldsymbol{\tau}_c$, $\bsym{\nu}_c$
\begin{align}
    &\mathbf{f}_r(\mathbf{r}) =  \sum_{p=0}^{P-1}\sum_{n=-N}^{N}[\mathbf{q}]_{v}\mathbf{G}_{v}^H \mathbf{a}(\mathbf{r}) \in \mathbb{C}^{J}\\&\mathbf{f}_c(\mathbf{c})=\sum_{p=0}^{P-1}\sum_{n=-N}^{N}[\mathbf{q}]_{v} \mathbf{A}^H_v \mathbf{a}(\mathbf{c}) \in \mathbb{C}^{PJ}
\end{align}
with complex coefficients $\mathbf{q}$ such that
\begin{align}
    \mathbf{f}_r(\mathbf{r}_\ell) &= \mathrm{sign}( [\bsym{\alpha}_r]_\ell) \mathbf{u} \hspace{1em} \text{if} \hspace{1em} \forall \mathbf{r}_\ell \in \mathcal{R} \label{eq:cert_1}\\
    \mathbf{f}_c(\mathbf{c}_q) &= \mathrm{sign}( [\bsym{\alpha}_c]_q) \mathbf{v} \hspace{1em} \text{if} \hspace{1em} \forall\mathbf{c}_q \in \mathcal{C}  \label{eq:cert_2}\\
    \Vert \mathbf{f}_r(\mathbf{r}) \Vert_2^2 &< 1 \hspace{1em } \forall \mathbf{r} \in [0,1]^2 \setminus \mathcal{R}  \label{eq:cert_3}\\
    \Vert \mathbf{f}_c(\mathbf{c}) \Vert_2^2  &< 1 \hspace{1em }  \forall \mathbf{c} \in [0,1]^2 \setminus \mathcal{C}\label{eq:cert_4}
\end{align}
where $\operatorname{sign}(c) = \frac{c}{|c|}$ .
\begin{proof}
The variable $\mathbf{q}$ is dual feasible. \par\noindent\small
%\langle\mathbf{q,\aleph_r(\mathbf{Z}_r)}\rangle_{\mathbb{R}}+ \langle\aleph_c(\mathbf{Z}_c)\rangle_{\mathbb{R}}
\begin{flalign}
    &\langle\mathbf{q,y}\rangle_{\mathbb{R}} = \langle\aleph_r^*(\mathbf{q}),\mathbf{Z}_r\rangle_{\mathbb{R}}+ \langle\aleph_c^*(\mathbf{q}),\mathbf{Z}_c\rangle_{\mathbb{R}}\nonumber\\
    %& =\langle\aleph_r^*(\mathbf{q}),\sum_{\ell=0}^{L-1} [\bsym{\alpha}_r]_\ell\mathbf{u}\mathbf{a}(\mathbf{r}_\ell)^H\rangle_{\mathbb{R}} + \langle\aleph_c^*(\mathbf{q}),\sum_{q=0}^{Q-1} [\bsym{\alpha}_c]_q\mathbf{v}\mathbf{a}(\mathbf{c}_q)^H\rangle_{\mathbb{R}} \nonumber \\
    & = \sum_{\ell=0}^{L-1} [\bsym{\alpha}_r]_\ell^* \langle\aleph_r^*(\mathbf{q}),\mathbf{u}\mathbf{a}(\mathbf{r}_\ell)^H\rangle_{\mathbb{R}} + \sum_{q=0}^{Q-1} [\bsym{\alpha}_c]_q^* \langle\aleph_c^*(\mathbf{q}),\mathbf{v}\mathbf{a}(\mathbf{c}_q)^H\rangle_{\mathbb{R}}\nonumber \\
    &= \sum_{\ell=0}^{L-1} [\bsym{\alpha}_r]_\ell^* \langle\mathbf{f}_r(\mathbf{r}_\ell),\mathbf{u}\rangle_{\mathbb{R}} + \sum_{q=0}^{Q-1} [\bsym{\alpha}_c]_q^* \langle\mathbf{f}_c(\mathbf{c}_q),\mathbf{v}\rangle_{\mathbb{R}}\nonumber\\
    & =\sum_{\ell=0}^{L-1}[\bsym{\alpha}_r]_\ell^*\textrm{sign}( [\bsym{\alpha}_r]_\ell) + \sum_{q=0}^{Q-1} [\bsym{\alpha}_c]_q^*\textrm{sign}( [\bsym{\alpha}_c]_q)\nonumber\\
    &= \sum_{\ell=0}^{L-1} |[\bsym{\alpha}_r]_\ell| + \sum_{q=0}^{Q-1} |[\bsym{\alpha}_c]_q|\geq ||\mathbf{Z}_r||_{\mathcal{A}_r} + ||\mathbf{Z}_c||_{\mathcal{A}_c}.\label{eq:lower_bound_dual}
\end{flalign}\normalsize
On the other hand, it follows from H\"{o}lder inequality that
\begin{align}
\langle\mathbf{q,y}\rangle_{\mathbb{R}}&=\langle\aleph_r^*(\mathbf{q}),\mathbf{Z}_r\rangle_{\mathbb{R}}+ \langle\aleph_c^*(\mathbf{q}),\mathbf{Z}_c\rangle_{\mathbb{R}}\\
&\leq ||\aleph_r^*(\mathbf{q})||_{\mathcal{A}_r}^*||\mathbf{Z}_r||_{\mathcal{A}_r} +  \leq ||\aleph_c^*(\mathbf{q})||_{\mathcal{A}_c}^*||\mathbf{Z}_c||_{\mathcal{A}_c}\\
&\leq ||\mathbf{Z}_r||_{\mathcal{A}_r}+||\mathbf{Z}_c||_{\mathcal{A}_c},
\label{eq:upper_bound_dual}
\end{align}
where the first inequality is due to Cauchy-Schwarz inequality and the last inequality follows from \eqref{eq:cert_1}, \eqref{eq:cert_2}, \eqref{eq:cert_3}, and \eqref{eq:cert_4}. Therefore, based on \eqref{eq:lower_bound_dual} and \eqref{eq:upper_bound_dual}, we conclude that $\langle\mathbf{q,y}\rangle_{\mathbb{R}}=||\mathbf{Z}_r||_{\mathcal{A}_r}+||\mathbf{Z}_c||_{\mathcal{A}_c}$ showing that the pair $(\mathbf{Z}_r,\mathbf{Z}_c)$ is primal optimal and, from strong duality, $\mathbf{q}$ is dual optimal. 
%\vspace{-1em}
\end{proof}
\end{proposition}
The existence of the polynomial has been previously demonstrated in 2-D (non-blind) deconvolution of a radar channel \cite{heckel2016super} as well as in the blind case \cite{suliman2021mathematical}. In our problem, two different trigonometric polynomials are directly related through the dual variable $\mathbf{q}$. Thus, to prove the existence of these polynomials under this constraint implies a more extended analysis that the above-mentioned works \cite{heckel2016super}, \cite{suliman2021mathematical}. Finally, even though this paper focuses on the channel estimation, the information embedded in the atoms $\mathbf{Z}_r$ and $\mathbf{Z}_c$ also allow for exact estimation of the radar waveform and the communications symbols. We omit the proof of the existence of these polynomials and the guarantees for the unique solution of $\mathbf{Z}_r$ and $\mathbf{Z}_c$ because of paucity of space. Briefly, the proof follows from \cite{candes_superresolution ,yang2016super, chi2016guaranteed}, where the polynomials are formulated as linear combination of fast decaying random kernels (\textit{e.g.} randomized F\'ejer kernel) and their derivatives with the constraint that the polynomials have common kernels.% such that, under some conditions, isotropy and incoherence properties of the subspace representation \cite{chi2016guaranteed}. they achieve  \eqref{eq:cert_1},\eqref{eq:cert_2},\eqref{eq:cert_3},\eqref{eq:cert_4}\cite{heckel2016super}.
\section{Numerical Experiments}%\vspace{-0.2cm}
\label{sec:results}
To numerically validate the proposed method, we set $M = 13, P=9, Q=L=3$ and $J=3$. The delay-Doppler parameters were taken from a random uniform distribution, which results in $\bsym{\tau}_r = [0.23, 0.68,0.87]$, $\bsym{\nu}_r = [0.45, 0.42,0.71]$, $\bsym{\tau}_c = [0.12, 0.21,0.95]$, and $\bsym{\nu}_c = [0.09, 0.25,0.87] $. The columns of the transformation matrices $\mathbf{B}$ and $\mathbf{D}_p$ were generated following the distribution described in \cite{chi2016guaranteed}, i.e. $\mathbf{b}_n= [1, e^{\mathrm{j}2\pi \sigma_n}, \dots, e^{\mathrm{j}2\pi(J-1) \sigma_n}]$, where $\sigma_n \sim \mathcal{N}(0,1)$. The parameters $\bsym{\alpha}_r$ and $\bsym{\alpha}_c$ are drawn from a normal distribution with $|[\bsym{\alpha}_r]_\ell|=|[\bsym{\alpha}_c]_q|=1 \;\forall q, \forall \ell$. The coefficient vectors $\mathbf{u}, \mathbf{v}$ are generated from a normal random distribution and normalized $\Vert\mathbf{u}\Vert=\Vert\mathbf{v}\Vert=1$. 

\begin{figure}[t]
    \centering
    \includegraphics[width=0.75
    \linewidth]{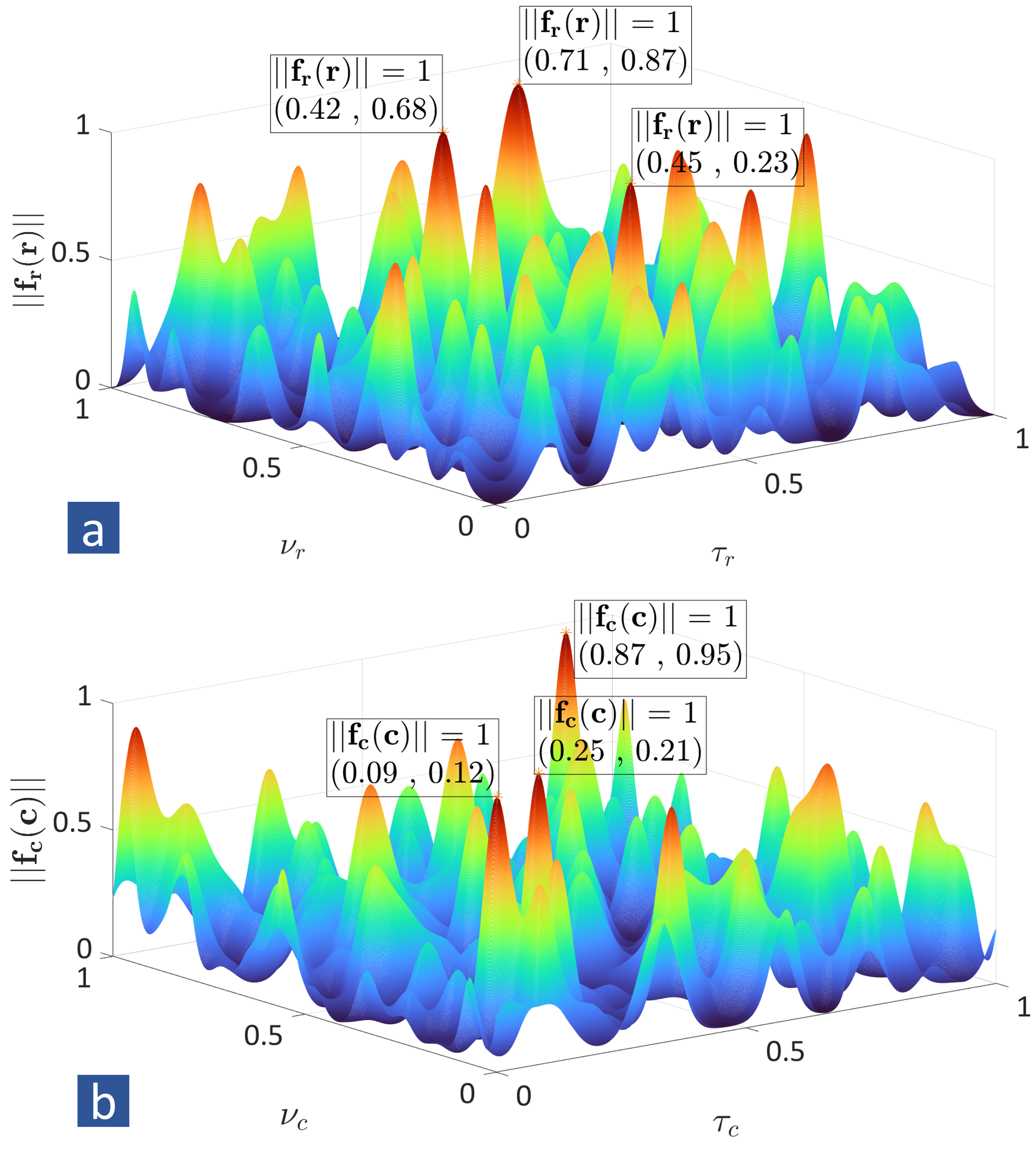}
    \caption{Dual polynomials corresponding to estimating the (a) radar and (b) communications channels in dual-blind deconvolution set-up. The locations of the parameters are given when the maximum modulus of the polynomials is unity.}
    \label{fig:results_dual}
\end{figure}

We solve the SDP optimization problem in \eqref{dual_opt} using CVX SDP3 solver \cite{grant2009cvx}. With the dual solution, we build the dual trigonometric polynomials by evaluating them in a discrete 2-D time-delay and doppler domain with a sampling step of $1e-3$. The resulting dual polynomials are depicted in Fig. \ref{fig:results_dual}, showing that the set of the radar and communications channels parameter are exactly located when $\Vert\mathbf{f}_r(\mathbf{r})\Vert = 1$ and $\Vert\mathbf{f}_c(\mathbf{c})\Vert = 1$. 

\begin{figure}[t]
    \centering
    \includegraphics[width=0.95\linewidth]{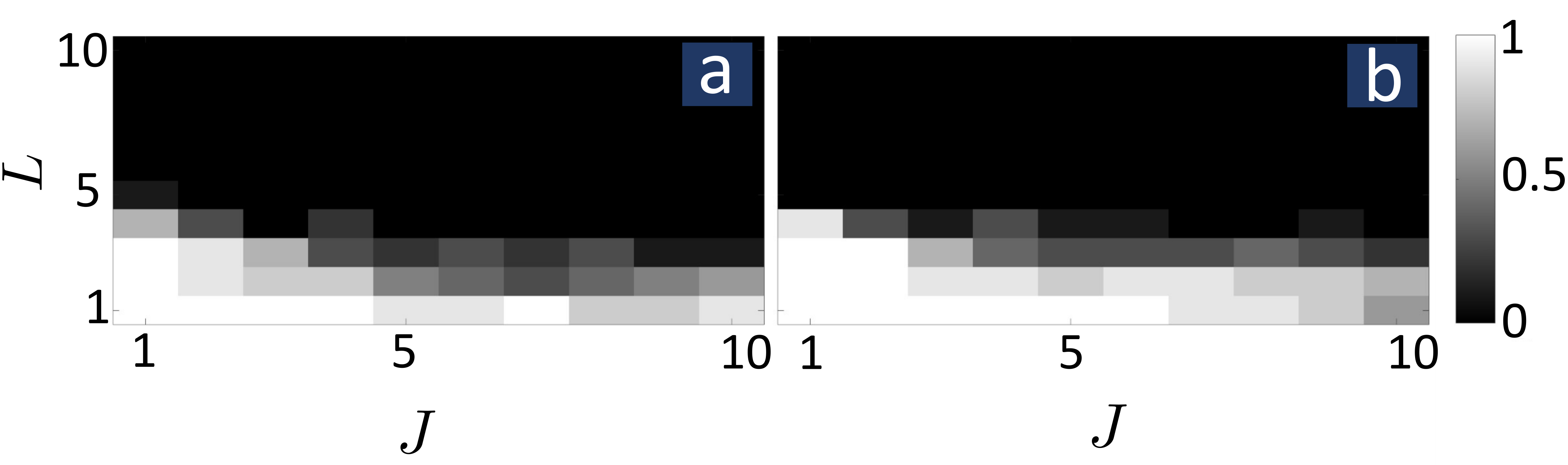}%\vspace{-0.5em}
    \caption{Exact localization probability of (a) radar and (b) communications channels as a function of the number of the targets and paths $Q=L$ and the subspace size $J$. }
    \label{fig:stats}%\vspace{-2em}
\end{figure}

Fig. \ref{fig:stats} shows the phase transition of the proposed method by varying the number of the targets and paths $Q=L$ and the subspace size $J$ for $10$ realizations of $L$ and $J$. The other parameters are same as before. We declare a successful estimation when $\Vert \mathbf{r} - \hat{\mathbf{r}}\Vert_2<1e^{-3}$ and $\Vert \mathbf{c} - \hat{\mathbf{c}}\Vert_2<1e^{-3}$. 
%\vspace{-0.5cm}
\section{Summary}
\label{sec:summ}
%\vspace{-0.3cm}
We proposed a dual-blind deconvolution method for jointly estimating the channel parameters of radar and communications systems from the information acquired with a single receiver. The proposed approach leverage the sparse structure of the channels to formulate the recovery problem as the minimization of two atomic norms corresponding to the radar and communications signals. The dual problem leads to the construction of two trigonometric polynomials corresponding to the radar and communications signal directly coupled through the dual variable. This problem is reformulated as an equivalent SDP by harnessing the parametrization of the dual trigonometric polynomials using positive semidefinite matrix and efficiently solved using off-the-shelf solvers. Numerical experiments validate the proposed approach that estimates the radar and communications channels parameters perfectly. 

\clearpage
\bibliographystyle{IEEEtr}
\bibliography{biblio.bib}

\end{document}